\documentclass[preprint,aps]{revtex4}
\usepackage{mathrsfs}

\usepackage{graphicx}
\usepackage{multirow}
\usepackage{makecell}
\usepackage{booktabs}

\usepackage{dcolumn}
\usepackage{bm}
\usepackage{xcolor}

\newcommand{\ef}{$E_{\rm F}$}

\newcommand{\dz}{$d_{z^2}$}

\newcommand{\dxy}{$d_{xy}$}

\newcommand{\rev}[1]{\textcolor{black}{#1}}

\begin{document}



\title{Correlation-Driven Electronic Reconstruction in FeTe$_{1-x}$Se$_x$}

\author{Jianwei Huang$^{1}$, Rong Yu$^{2}$, Zhijun Xu$^{3,4,5}$, Jian-Xin Zhu$^{6,7}$, Ji Seop Oh$^{5,1}$, Qianni Jiang$^{8}$, Meng Wang$^{9}$, Han Wu$^{1}$, Tong Chen$^{1}$, Jonathan D. Denlinger$^{10}$, Sung-Kwan Mo$^{10}$, Makoto Hashimoto$^{11}$, Matteo Michiardi$^{12,13,14}$, Tor M. Pedersen$^{15}$, Sergey Gorovikov$^{15}$, Sergey Zhdanovich$^{12,13}$, Andrea Damascelli$^{12,13}$, Genda Gu$^{16}$, Pengcheng Dai$^{1}$, Jiun-Haw Chu$^{8}$, Donghui Lu$^{11}$, Qimiao Si$^{1}$, Robert J. Birgeneau$^{5,17,18,*}$, Ming Yi$^{1,5,*}$}
\affiliation{
\\$^{1}$Department of Physics and Astronomy, Rice Center for Quantum Materials, Rice University, Houston, Texas 77005, USA
\\$^{2}$Department of Physics, Renmin University of China, Beijing 100872, China
\\$^{3}$NIST Center for Neutron Research, National Institute of Standards and Technology, Gaithersburg Maryland 20899, USA
\\$^{4}$Department of Materials Science and Engineering, University of Maryland, College Park, Maryland 20742, USA
\\$^{5}$Department of Physics, University of California Berkeley, Berkeley, California 94720, USA
\\$^{6}$Theoretical Division, Los Alamos National Laboratory, Los Alamos, New Mexico 87545, USA
\\$^{7}$Center for Integrated Nanotechnologies, Los Alamos National Laboratory, Los Alamos, New Mexico 87545, USA
\\$^{8}$Department of Physics, University of Washington, Seattle, Washington 98195, USA
\\$^{9}$School of Physics, Sun Yat-sen University, Guangzhou, Guangdong 510275, China
\\$^{10}$Advanced Light Source, Lawrence Berkeley National Lab, Berkeley, California 94720, USA
\\$^{11}$Stanford Synchrotron Radiation Lightsource, SLAC National Accelerator Laboratory, Menlo Park, California 94025, USA
\\$^{12}$Department of Physics $\&$ Astronomy, University of British Columbia, Vancouver BC V6T 1Z1, Canada
\\$^{13}$Quantum Matter Institute, University of British Columbia, Vancouver BC V6T 1Z4, Canada
\\$^{14}$Max Planck Institute for Chemical Physics of Solids, N\"{o}thnitzer Straße 40, 01187 Dresden, Germany
\\$^{15}$Canadian Light Source, Inc., 44 Innovation Boulevard, Saskatoon SK S7N 2V3, Canada
\\$^{16}$Condensed Matter Physics and Materials Science Department, Brookhaven National Laboratory, Upton, New York 11973, USA
\\$^{17}$Materials Sciences Division, Lawrence Berkeley National Laboratory, Berkeley, California 94720, USA
\\$^{18}$Department of Materials Science and Engineering, University of California Berkeley, Berkeley, California 94720, USA
\\$^{*}$To whom correspondence should be addressed: mingyi@rice.edu and robertjb@berkeley.edu
}

\date{\today}

\begin{abstract}
Electronic correlation is of fundamental importance to high temperature superconductivity. While the low energy electronic states in cuprates are dominantly affected by correlation effects across the phase diagram, observation of correlation-driven changes in fermiology amongst the iron-based superconductors remains rare. Here we present experimental evidence for a correlation-driven reconstruction of the Fermi surface tuned independently by two orthogonal axes of temperature and Se/Te ratio in the iron chalcogenide family FeTe$_{1-x}$Se$_x$. We demonstrate that this reconstruction is driven by the de-hybridization of a strongly renormalized $d_{xy}$ orbital with the remaining itinerant iron 3$d$ orbitals in the emergence of an orbital-selective Mott phase. Our observations are further supported by our theoretical calculations to be salient spectroscopic signatures of such a non-thermal evolution from a strongly correlated metallic phase into an orbital-selective Mott phase in $d_{xy}$ as Se concentration is reduced.

\end{abstract}

\maketitle

\newpage

\section{Introduction}
The phase diagrams of the copper oxide superconductors and iron-based superconductors (FeSC) have often been compared to demonstrate a degree of similarity---a superconducting dome emerging from the suppression of a magnetic ground state~\cite{Paglione2010, Wang2011, Keimer2015}. Electron correlations as well as effects of symmetry-breaking electronic orders have been observed in both~\cite{Lee2006, Keimer2015, Yin2011, Yi2017a}. However, the dominant mechanism that drives the evolution of the Fermi surface and low energy quasiparticles differs in the two material systems. In the cuprates, the parent compound is an antiferromagnetic (AFM) Mott insulator, from which low energy quasiparticles emerge as the system is doped away from half-filling, eventually approaching a Fermi liquid on the overdoped side~\cite{Lee2006, Chen2019a}. In most FeSCs, the parent compound is a metal harboring a spin-density wave order~\cite{DeLaCruz2008}, where the Fermi surface is well-defined and can be largely understood as a modified version of the non-magnetic fermiology reconstructed by the symmetry-breaking electronic orders~\cite{Yi2011a}. Therefore, while arguably the dominant mechanism that controls the evolution of the fermiology in the cuprates is electron correlation effects, that for the FeSCs have largely been understood to be the symmetry-breaking electronic orders~\cite{Yi2017a, Sobota2021}.

However, it has also been recognized early on that the FeSCs exhibit moderate correlations~\cite{Yin2011}. The analysis of the experimentally observed Drude weight from optical conductivity puts the FeSCs between the limits of the Mott insulator and the itinerant metal~\cite{Qazilbash2009}. Moreover, from both theory and experimental observation via angle-resolved photoemissoin spectroscopy (ARPES), it has been recognized that such correlations are orbital-dependent~\cite{Yin2011,DeMedici2014,Si2016,Yi2017a}. In particular, while the FeSCs as bad metals are deemed to be moderately correlated compared to the cuprate high temperature superconductors, the electrons belonging to different orbitals are found to be correlated to different degrees -- stronger in the $d_{xy}$ orbital than in the $d_{xz}$/$d_{yz}$ orbitals~\cite{Yi2017a}. This orbital-differentiation is enhanced by Hund’s coupling, and increases systematically in sync with the vertical elongation of the iron tetrahedron from the iron-phosphides to iron-pnictides to iron-chalcogenides~\cite{Yi2017a}, in which strong orbital-selectivity has been reported~\cite{Yin2012,Yi2015,Pu2016}. From the electronic degree of freedom, the effective mass of the $d_{xy}$ orbital dominated band has been reported to be larger than that of the $d_{xz}$/$d_{yz}$ orbitals. From both the nuclear magnetic resonance and neutron scattering measurements, coexistence of both itinerant and local spins has been found~\cite{Dai2012}, where the $d_{xy}$ orbital contributes dominantly to the local spin susceptibility~\cite{Li2020,Li2018,Song2018}. 
As an example, inelastic neutron scattering experiments on detwinned NaFeAs, a parent compound of FeSCs, have shown that spin waves of the system are orbital-selective with high energy spin waves arising mostly from the $d_{xy}$ orbital and obeying the C$_4$ rotational symmetry, while the low energy spin waves are from the $d_{xz}$/$d_{yz}$ orbitals that break the C$_4$ rotational symmetry below the nematic phase transition temperature~\cite{Tam2020}.  

A question then arises---while the fermiology of the FeSCs is far from that of a Mott insulator, could the strong localization of a selective orbital be sufficient to induce significant modifications to the fermiology and associated low energy electronic states? 

Generically in a multiorbital material, this can be realized in an orbital-selective Mott phase (OSMP), where selected orbitals can become completely localized and gapped out from the Fermi level while other orbitals maintain a degree of itinerancy~\cite{DeMedici2005,Ferrero2009,Yin2011,Greger2013,Yu2013a,Georges2013,DeMedici2014}. As the system evolves into an OSMP, the hybridization between the electrons belonging to the strongly localized orbitals and the itinerant orbitals gradually turns off, leading to a reconstruction of the fermiology. In the larger context of quantum materials, evidence for OSMP has been reported for Ca$_{2-x}$Sr$_x$RuO$_4$~\cite{Anisimov2002,Neupane2009,Kim2021}, VO$_2$~\cite{Mukherjee2016}, and a transition metal dichalcogenide~\cite{Qiao2017}, in the form of strongly orbital-selective mass enhancement and spectral weight gapped out at the Fermi level~\cite{Neupane2009, Shimoyamada2009, Saeki2009, Ang2013, Sutter2017, Muraoka2018}. \rev{However, continuous evolution of Fermi surface reconstruction across an OSMP has been challenging to show for any material system, and has only recently been reported for Ca$_{2-x}$Sr$_x$RuO$_4$~\cite{Kim2021}.}

In this context, the strongly correlated iron-chalcogenide materials provide an opportunity to realize such correlation-driven Fermi surface reconstruction. To examine this, we consider the regime where evidence for a thermal crossover into an OSMP has recently been reported in FeTe$_{1-x}$Se$_x$~\cite{Yi2013, Yi2015}. In these prior works, the $d_{xy}$ orbital is strongly renormalized at low temperatures and loses spectral weight as temperature is raised sufficiently high~\cite{Yi2013, Yi2015}. To examine the effect of such strong correlations on the fermiology, it is desired to go beyond the study of any thermal crossover and, instead, scan the ground state landscape by tuning a non-thermal control parameter while staying at low temperatures. An added benefit of doing the latter is that the Fermi surface is only sharply defined at low temperatures, and studying the quantum phase transition through a non-thermal parameter variation allows for unambiguously detection of Fermi surface reconstruction. 

FeTe$_{1-x}$Se$_x$ is a prototypical iron-based superconductor with the simplest crystal structure consisting of iron-chalcogen layers~\cite{Hsu2008}. Recent studies have also revealed the existence of topological surface states and potential Majorana zero modes in this material platform~\cite{Zhang2018,Wang2018,Li2021}. \rev{The parent compound, FeTe, is an AFM metal~\cite{Fruchart1975,Bao2009,Li2009,Liu2010}. With the isovalent substitution of Se on Te sites, superconductivity emerges, reaching a maximum $T_{\rm c}$ of 14.5 K in FeTe$_{0.56}$Se$_{0.44}$~\cite{Liu2010,Martinelli2010}. It is known that as-grown single crystals of FeTe$_{1-x}$Se$_x$ have a tendency to harbor interstitial excess iron, which leads to spin-glass behavior and incoherence in the low energy electronic spectra~\cite{Katayama2010, Ieki2014}. It has been shown that the excess Fe can be reduced or completely removed by annealing in either an oxygen or Te-vapor environment~\cite{Sun2015, Dong2011, Lin2015, Sun2015a}, which suppresses the spin-glass behavior and results in a phase diagram that bares a closer resemblance to that of iron pnictides~\cite{Sun2016}.} 

In this work, we present systematic evidence for a reconstruction of the Fermi surface driven by orbital-dependent correlation effects in the FeTe$_{1-x}$Se$_x$ family of superconductors as the phase diagram is traversed non-thermally along the Se-substitution axis as well as along the temperature axis. Remarkably, strong reorganization of the Fermi surface is observed in the absence of any symmetry-breaking electronic order. We demonstrate how the more coherent $d$-orbitals, in particular, the $d_{z^2}$ orbital, serve as a transparent means to visualize the suppression of the $d_{xy}$ states at the Fermi energy by gradually appearing at the Fermi level as its hybridization with $d_{xy}$ is turned off. In addition, we show that the Fermi surface reconstruction is accompanied by an orbital-dependent mass enhancement and signatures of a lower Hubbard band as we tune from FeSe towards FeTe. Combined with the temperature axis, we not only arrive at a comprehensive phase diagram of the orbital-selectivity in FeTe$_{1-x}$Se$_x$, the understanding of which is discussed in connection with anomalies reported in measurements of the Hall coefficient, resistivity, and magnetic excitations, but also achieve a direct $k$-space visualization of the correlation-driven reconstruction of the electronic states near the Fermi energy.

\section{Results}
\subsection{Fermi surface reconstruction}
\begin{figure}
\includegraphics[width=0.55\textwidth]{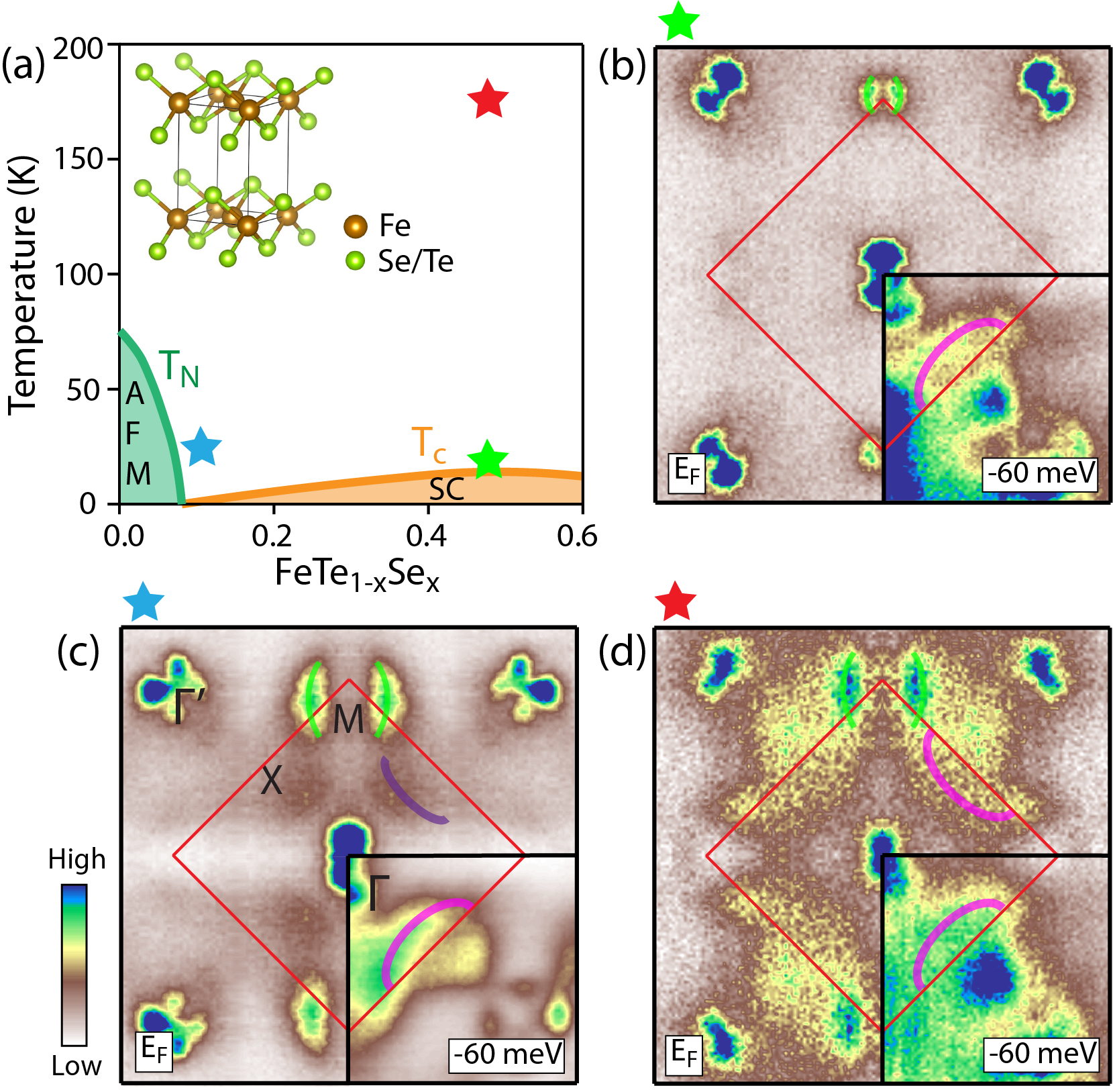}
\caption{\label{fig:Fig1}Fermi surface evolution across the phase diagram of FeTe$_{1-x}$Se$_{x}$. (a) Phase diagram of FeTe$_{1-x}$Se$_{x}$ showing where the Fermi surfaces are measured. (b) Fermi surface and constant energy contour measured at -60 meV (lower right inset) for FTS44 taken at 18 K. (c)-(d) Same as (b) but taken on FTS11 at 24 K and FTS44 at 174 K, respectively. The red diamonds mark the 2-Fe Brillouin zone.
}
\end{figure}

We begin by presenting a summary of our key results on the evolution of the Fermi surface along both the temperature and Se substitution axes of the FeTe$_{1-x}$Se$_x$ phase diagram (Fig.~\ref{fig:Fig1}a). For simplicity, we refer to single crystals with the following Se content: \textit{x} = 0, 0.11, 0.19, 0.28, 0.44 as FT, FTS11, FTS19, FTS28, FTS44, respectively. Starting from the familiar low temperature state of the optimally-substituted FTS44 (Fig.~\ref{fig:Fig1}b), the Fermi surface dramatically changes compared to both the low temperature Fermi surface of FTS11 just outside the magnetic ground state near FeTe (Fig.~\ref{fig:Fig1}c) and its own high temperature state above the crossover into the OSMP (Fig.~\ref{fig:Fig1}d) \cite{Yi2015}. We note that in the region of the phase diagram spanning these three measurement points (Fig.~\ref{fig:Fig1}a), there exists no known symmetry-breaking electronic orders. Yet, obvious changes in the Fermi surface can be observed. In low temperature FTS44, as previously reported~\cite{Chen2010, Tamai2010, Nakayama2010, Miao2012, Okazaki2012b, Lubashevsky2012, Liu2015a}, the Fermi surface consists of small hole pockets near the Brillouin zone (BZ) center ($\Gamma$) and electron pockets near the BZ corner (M). The wavevector difference between them matches the momentum transfer, where neutron spin resonance has been observed in the superconducting state~\cite{Qiu2009}. In both the high temperature state of FTS44 and low temperature state of FTS11, the apparent fermiology has drastically changed in a similar fashion. The electron pocket at the M point is expanded while a new dominant arc-like spectral weight (marked in magenta and purple) appears along the BZ boundary centered at the X point. Interestingly, this arc feature is already present in the constant energy contours at -60 meV for all three measurement points, in contrast to the Fermi surface. This observation suggests that the modifications along the substitution and temperature axes may have a similar origin. To understand this fermiology change, we examine the evolution across each axis of the phase diagram in turn.

\begin{figure*}
\includegraphics[width=0.98\textwidth]{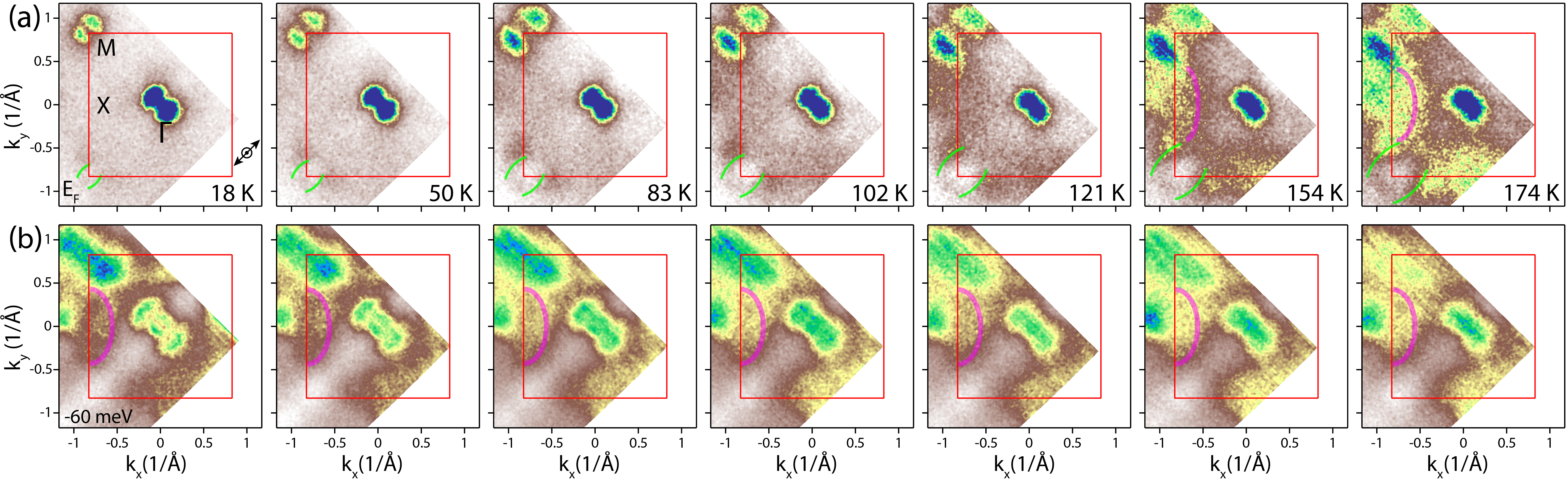}
\caption{\label{fig:Fig2}Fermi surface evolution with temperature for FTS44. (a) Measured Fermi surfaces as temperature is varied between 18 K and 174 K. (b) Simultaneously measured constant energy contours at -60 meV showing no change as a function of temperature. Polarization is as marked in the left-most panel of (a) indicating an in-plane polarization along $\Gamma$-M with a finite out-of-plane component. The red squares mark the Brillouin zone of the 2-Fe unit cell.
}
\end{figure*}

We first present the measured Fermi surface of optimal FTS44 taken as a function of temperature from 18 K up to 174 K (Fig.~\ref{fig:Fig2}a). Here we see the smooth evolution of the two changes previously mentioned: i) expansion of the electron pockets at the M point, and ii) emergence of an arc-like feature near the BZ boundary along X. While the Fermi surface is changing, we note that at -60 meV, no change is observed through this entire temperature range, where the spectral weight around X already appears at this binding energy even at the lowest temperature (Fig.~\ref{fig:Fig2}b). This contrast suggests that the new feature observed near X at \ef~emerges from higher binding energy when temperature is raised.

Next, we present the measured Fermi surface of FeTe$_{1-x}$Se$_x$ along the substitution-axis at low temperatures across the phase diagram (Fig.~\ref{fig:Fig3}). To avoid complexity due to Fermi surface reconstruction from the bicollinear AFM phase of FT, we use the Fermi surface measured above the magnetic ordering temperature for FT in this comparison (Fig.~\ref{fig:Fig3}a, the AFM state in FT is addressed in Supplementary Note 1). Hence across this comparison, there is no symmetry-breaking order. Yet, the same two qualitative changes can be observed with the decrease of the Se ratio (\textit{x}): i) expansion of the electron pockets near the M point, and ii) emergence of the arc-like feature near the X point. As has been shown for the temperature evolution, the changes are limited to a small energy window near \ef, where the constant energy contours at -60 meV also show no obvious change with varying Se content (Fig.~\ref{fig:Fig3}b). Comparing the Fermi surface across different Se content levels, the better resolved arc-like feature appears to evolve from a faint intensity near $\Gamma$, which is the spectral weight of the $d_{xy}$ hole band in FTS44. This evolution is furthermore evident from the band dispersion measured along the $\Gamma$-X direction (Fig.~\ref{fig:Fig3}c). As can be seen from both the dispersion images as well as the momentum distribution curve (MDC) stacks, a hole-like band emanating from the $\Gamma$ point in FTS44 shifts toward the X point with decreasing \textit{x}, eventually becoming a nearly vertical dispersion, giving rise to the arc-like feature that appears at \ef~in FT. This also explains the observation that this arc-like feature is present already at high binding energy (-60 meV) across the different \textit{x} presented in the phase diagram. From a set of detailed polarization-dependence measurements, the origin of this arc-like feature is identified to be predominantly of \dz~orbital (see Supplementary Note 2 for photoemission matrix element analysis), which gradually evolves from high binding energy to \ef~with decreasing \textit{x} (Fig.~\ref{fig:Fig3}d, e and f). We will discuss the spectral weight evolution later.

\subsection{Orbital-dependent mass enhancement}
The striking modifications to the low energy electronic states and fermiology via both temperature and the non-thermal substitution-axis in absence of any symmetry breaking phases calls for an examination of the electronic correlations in this phase diagram. To determine the orbital-dependent correlation strength, we focus on the low-temperature (10 K) band renormalizations across the phase diagram near the BZ center. Along the high symmetry direction $\Gamma$-M, three hole-like bands are observed near \ef~for all Se content which are better visualized from the second derivative plots (Fig.~\ref{fig:Fig3}g). Consistent with previous results~\cite{Nakayama2010}, these three bands from the innermost to the outermost are identified as dominantly $d_{xz}$, $d_{yz}$, and $d_{xy}$, respectively. As the phase diagram is traversed along the substitution-axis at 10 K, it is evident that the band curvature of the $d_{yz}$ band does not vary strongly while that of the $d_{xy}$ band flattens considerably with decreasing Se content. Since the effective mass for each band is proportional to the inverse of the band curvature, we can extract the orbital-dependent mass enhancement from a parabolic fitting of each band. The mass enhancement obtained from the ratio between the fitted effective mass $m^*$ to that of first-principle calculations $m_{\rm DFT}$ is plotted as a function of Se content in Fig.~\ref{fig:Fig3}h, consistent with the trend from previous results for higher Se content~\cite{Maletz2014,Liu2015a}. First, the mass enhancement of $40$ is unusually large for 3$d$-electrons near the occupancy $n=6$, even for the iron chalcogenides. Second, we observe an orbital-dependent band renormalization. The $d_{xy}$ orbital has a much larger band renormalization factor than the $d_{yz}$ orbital. Third, while the mass enhancement of the $d_{yz}$ orbital rises slowly with decreasing Se ratio \textit{x}, a divergent behavior of the mass enhancement for the $d_{xy}$ orbital is observed with decreasing Se content. The inverse of the $d_{xy}$ mass enhancement shows a linear trend versus Se ratio (Fig.~\ref{fig:Fig3}h). These results strongly indicate that FeTe is in proximity to an OSMP ground state and further suggest that the evolution of the apparent fermiology is associated with the gradual disappearance of the $d_{xy}$ spectral weight as \textit{x} is reduced.

\subsection{Visualizing the selective localization of $d_{xy}$-electrons via $d_{z^2}$-electrons}
The orbital-dependent band renormalization, diverging $d_{xy}$ effective mass, and modification of the Fermi surface, taken together, can be understood as manifestations of the tendency of the FeTe$_{1-x}$Se$_x$ system towards an OSMP \rev{with decreasing x. On the verge of Mott localization, the disappearance of the $d_{xy}$ spectral weight near \ef~results in a redistribution of the residual electronic states at \ef~as the FeTe end of the phase diagram is approached. How does the diminishing $d_{xy}$ weight at the Fermi level lead to the crossing of the $d_{z^2}$ band through the Fermi level? Qualitatively, we attribute this to the hybridization picture, which is believed to be important to the OSMP transition in the iron-based systems~\cite{Yu2017}.} As symmetry dictates a nonzero hybridization matrix between the $d_{xy}$ and $d_{z^2}$ states near X, they are allowed to hybridize. Analogous to the appearance of a ``large'' Fermi surface in the $f$-electron heavy fermion metals, the itinerant $d_{xy}$ states in FeTe$_{1-x}$Se$_x$ hybridize with the $d_{z^2}$ state near X in the Fermi surface formation. As a result, the hybridization gap removes the $d_{z^2}$ states from the Fermi level. However, just like the localization of the $f$-electrons in heavy fermion systems leads to a ``small'' Fermi surface of the $spd$ conduction electrons, localizing the $d_{xy}$-electron state allows the $d_{z^2}$-electron state to cross the Fermi level near X.

To substantiate this qualitative picture, we have carried out calculations based on a five-orbital Hubbard model for FeTe$_{1-x}$Se$_x$ using the U(1) slave-spin method (for the method and the details of the calculation, see Supplementary Note 3 and 4). The substitution of larger Te atoms for smaller Se atoms increases the Fe-Se/Te bond length and decreases the Fe-Te/Se-Fe bond angle. The former effect increases the overall correlation strength by lengthening the dominant hopping path while the latter effect suppresses the effective hopping for the largely in-plane $d_{xy}$ orbital more than that of the other orbitals, pushing the system toward an OSMP\cite{Yi2017a}. 

\begin{figure*}
\includegraphics[width=0.98\textwidth]{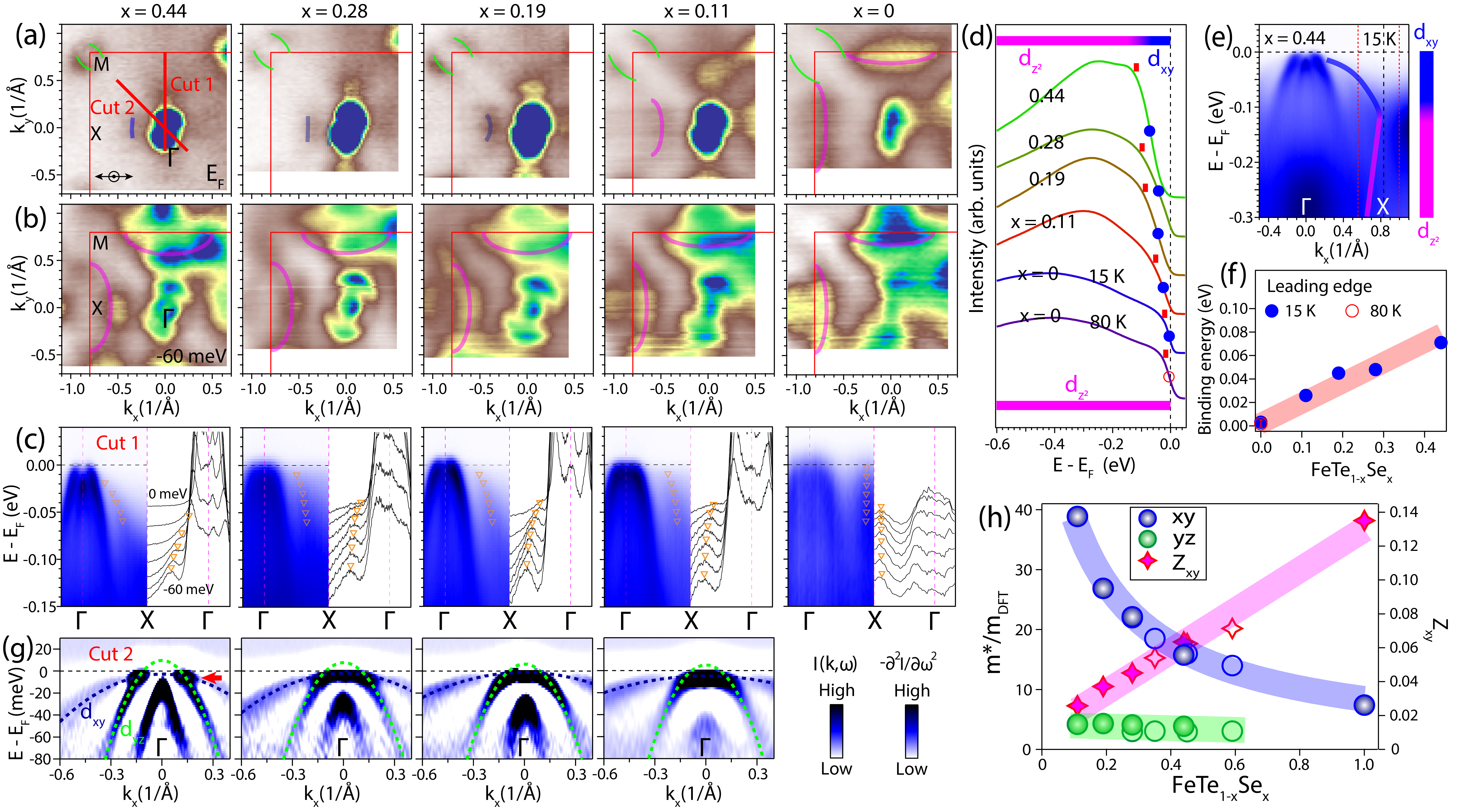}
\caption{\label{fig:Fig3} Fermi surface evolution with Se ratio $\textit{x}$. (a) Measured Fermi surfaces for different Se ratio \textit{x} at 15 K except that for \textit{x} = 0, which is measured at 80 K above its magnetic transition. (b) Simultaneously measured constant energy contours at -60 meV showing no change. (c) Corresponding band dispersions and momentum distribution curves (MDCs) along the $\Gamma$-X-$\Gamma$ direction. The markers are peaks fitted from MDCs. The MDCs are for binding energies from 0 meV to 60 meV. (d) Integrated energy distribution curve (EDC) around X of FeTe$_{1-x}$Se$_x$ with the integral range marked by the red dashed lines in (e). The energy positions where the integrated EDCs show a maximum curvature are marked by the red bars. The leading edge marked by the blue and red circles represents the energy position where the EDC intensity is half of where at the maximum curvature. (e) Band dispersion of FeTe$_{0.56}$Se$_{0.44}$ at 15 K along the $\Gamma$-X-$\Gamma$ direction. (f) Leading edge determined in (d) versus Se content of FeTe$_{1-x}$Se$_x$. (g) Band dispersions (second derivative) measured with 22 eV photons at 10 K along the $\Gamma$-M direction around the Brillouin zone center for different \textit{x}. Parabolic fittings of the $d_{yz}$ and $d_{xy}$ hole bands are overlaid. The red arrow in left-most panel of (g) points at the hybridization of $d_{yz}$ and $d_{xy}$ orbitals. (h) Extracted band enhancement factor $m^*$/$m_{\rm DFT}$ for $d_{yz}$ and $d_{xy}$ plotted as a function of \textit{x} represented by circles. The shaded lines are guides to the eyes. Diamond markers show the inverse of the $d_{xy}$ mass enhancement where the shaded pink line is a linear fit. The empty markers are reproduced from Liu \textit{et al}.~\cite{Liu2015a}. 
}
\end{figure*}

All salient features of our experimental observations presented are captured in this set of calculations, as shown by a direct comparison of the calculations for \textit{x} = 0.1 and \textit{x} = 0.5 (Fig.~\ref{fig:Fig4}). \rev{Since density-functional theory (DFT) calculations (see Supplementary Note 3) do not capture well the observed dispersions near \ef, a $U$ of $~$3 eV is used to be consistent with the experimentally observed band renormalizations (Supplementary Note 4)}. The projection onto the $d_{xy}$ and $d_{z^2}$ orbitals are shown in blue and magenta, respectively, for all calculations. In the strongly correlated metallic phase calculated for \textit{x} = 0.5 (Fig.~\ref{fig:Fig4}a), the $d_{xy}$ dominated bands are strongly renormalized. When projected onto a Fermi surface mapping within a finite integration energy window about \ef, the flattened $d_{xy}$ hole band intensity becomes largely incoherent, resulting in a faint large pocket spectra from the hole band around the $\Gamma$ point (Fig.~\ref{fig:Fig4}b). When the OSMP is entered, the band structure and Fermi surface are strongly reconstructed (Fig.~\ref{fig:Fig4}c and d). The origin of this reconstruction is most apparent along the $\Gamma$-X direction. In the strongly correlated metallic phase (Fig.~\ref{fig:Fig4}a), the strongly renormalized $d_{xy}$ bands interrupt a highly dispersive $d_{z^2}$ band near \ef, leading to very little $d_{z^2}$ spectral weight in the Fermi surface. In the OSMP (Fig.~\ref{fig:Fig4}c), the complete localization and gapping of the $d_{xy}$ orbital from \ef~frees the highly dispersive $d_{z^2}$ band, which now crosses \ef, forming the pocket around the X point (Fig.~\ref{fig:Fig4}d), consistent with our observed emergence of the arc-like feature around the X point. This fermiology is very consistent with that observed for the paramagnetic FeTe (right-most panel of Fig.~\ref{fig:Fig3}a). Regarding the intermediate evolution of the band dispersions, in the regime approaching the OSMP, the $d_{xy}$ bandwidth would narrow while its spectral weight diminishes, causing the hybridization between this $d_{xy}$ band and the highly dispersive d$_{z^2}$ band to decrease. This is consistent with our observed dispersions along the $\Gamma$-X direction, where the hole-like \dxy~band in the low temperature state of FTS44 gradually evolves to the nearly vertical dispersion as the \dz~band is recovered to \ef~(Fig.~\ref{fig:Fig3}c). 

We can experimentally track the disappearance of this hybridization as a function of Se ratio. In Fig.~\ref{fig:Fig3}d, we plot the integrated energy distribution curve (EDC) around the X point of FeTe$_{1-x}$Se$_x$ with different \textit{x} (Fig.~\ref{fig:Fig3}e). For FTS44, the integrated EDC shows a strong spectral weight suppression around \ef~due to the hybridization between the \dz~band and the \dxy~band (Fig.~\ref{fig:Fig3}e). Since the \dxy~band strongly disperses away from X towards $\Gamma$, it largely falls outside the integration window. The integrated EDC within this narrow momentum range is dominated by the \dz~spectral weight. The spectral suppression around \ef~is therefore an indirect way to estimate the bandwidth of the \dxy~band that hybridizes with the \dz~band (Fig.~\ref{fig:Fig3}d and e). As the \dxy~band is increasingly renormalized, the spectral suppression in the integrated EDC will decrease as the \dz~spectral weight fills towards \ef~until the suppression completely disappears with the localization of the \dxy~band in the OSMP. To capture the evolution of the spectral suppression with different \textit{x}, we identify the leading edge of the integrated EDCs for samples with different Se content (Fig.~\ref{fig:Fig3}d), and plot its energy as a function of Se content in Fig.~\ref{fig:Fig3}f. The leading edge approaches the Fermi level with decreasing Se content, demonstrating that the gapped-out \dz~spectral weight gradually fills in and forms the arc-like feature around X. Notably for FT, the leading edge shifts to \ef, indicating that it is in the OSMP at both 80 K and 15 K.

Besides the appearance of the arc-feature, the calculated $d_{xz}$/$d_{yz}$ electron pocket enlarges, as observed experimentally, to compensate the hole pocket formed by the nearly vertical band around X. As a function of \textit{x}, the calculated coherence factor, $Z$, for the $d_{xz}$/$d_{yz}$ orbital is compared to that of the $d_{xy}$ orbital (Fig.~\ref{fig:Fig4}e), where localization of the $d_{xy}$ orbital appears for \textit{x} $<$ 0.2 with $U$ $\sim$ 3 eV. \rev{We note that, the selective localization of the $d_{xy}$ states also causes the suppression of their hybridization with $d_{xz}$/$d_{yz}$ states.}
However, because the $d_{z^2}$-electron states are considerably more dispersive and gapped-out from \ef~due to the hybridization with $d_{xy}$, they serve as an especially transparent diagnostic of the OSMP in the way of Fermi surface reconstruction (Fig.~\ref{fig:Fig4}b, d).

\section{Discussion}
Finally, taking all the results together, we arrive at the phase diagram for the FeTe$_{1-x}$Se$_x$ material family (Fig.~\ref{fig:Fig5}). While previous results~\cite{Yi2015} indicate that for an optimally substituted compound at \textit{x} = 0.44, the OSMP can be reached via raised temperature at a characteristic temperature of 110 K, our observations reported here show that in the low temperature limit, the replacement of Se by Te also leads the material system towards the OSMP. This is supported by spectroscopic evidence of strongly orbital-selective band renormalization where the $d_{xy}$ effective mass tends toward a divergent behavior as the FeTe end is approached. Concomitantly, spectral weight from other orbitals redistribute near the \ef~to replace the diminishing $d_{xy}$ spectral weight. The characteristic crossover temperature for the OSMP therefore decreases with decreasing \textit{x}. This is confirmed by our measurement for a sample with \textit{x} = 0.25, where the temperature identified by the disappearance of $d_{xy}$ is measured to be 70 K (Supplementary Note 5). We note that considering the bandwidth of the $d_{xy}$ orbital ($\geq$ 30 meV, ~350 K), the observed disappearance of the $d_{xy}$ spectral weight as a function of Se ratio measured at 10 K cannot be caused by the thermal broadening effect, but rather due to the narrowing of the $d_{xy}$ bandwidth as the system enters into the OSMP.

Our understanding of the evolution from an orbital-dependent correlated metallic phase to an OSMP across the FeTe$_{1-x}$Se$_x$ phase diagram as observed from ARPES is also consistent with results reported by other probes. It has been reported that the compensated parent compounds of multi-band iron-based superconductors exhibit a negative Hall coefficient, $R_{\rm H}$, due to the dominance of the electron mobility. This is observed in isovalent-substituted BaFe$_2$(As,P)$_2$, where across the entire phase diagram, $R_{\rm H}$ remains negative~\cite{Fang2009,Kasahara2010}. For FeTe$_{1-x}$Se$_x$, however, there is a systematic change that varies as a function of substitution~\cite{Jiang2020}. This is clearly shown in the measurement over the phase diagram of Te-vapor treated FeTe$_{1-x}$Se$_x$ (Fig.~\ref{fig:Fig5})~\cite{Jiang2020}. Contrary to the behavior in BaFe$_2$(As,P)$_2$, Hall resistivity measurements on FeTe$_{1-x}$Se$_x$ exhibit a maximum and a subsequent sign-change at lower temperatures while the total charge carriers remain neutral~\cite{Jiang2020}. This turn-over behavior marked by the maximum indicates the onset of a competing behavior (Fig.~\ref{fig:Fig5}a). One interpretation of the negative Hall coefficient in the iron pnictides is the ($\pi$, $\pi$) spin fluctuations that make the hole carriers behave like electron carriers in the Hall measurements~\cite{Fanfarillo2012}. The localization of the $d_{xy}$ orbital would change the fermiology away from the nesting condition associated with the ($\pi$, $\pi$) spin fluctuations. This in turn suppresses the ($\pi$, $\pi$) spin fluctuations, which reduces the negative contribution to the Hall coefficient. This characteristic temperature scale extracted from $R_{\rm H}$ exhibits a similar trend with Se ratio \textit{x}, similar to $T_{\rm OSMT}$, which is the temperature where the spectral weight of $d_{xy}$ orbital vanishes with increasing temperature. 

We also note that in a recent magneto-transport measurement, such coherent-incoherent crossover is also reported~\cite{Otsuka2019}. It has also been pointed out that this coherent-incoherent crossover temperature scale of the $d_{xy}$ orbital is coupled to the $B_{2g}$ nematic susceptibility measured by elastoresistance, and hence suggests the potential important role played by the $d_{xy}$ orbital to nematicity in the iron-based superconductors~\cite{Jiang2020}. 

Previous studies of both Dynamic Mean Field Theory and ARPES have also identified the incoherent spectral weight in the high binding energy regime of FeSe as the lower Hubbard band (LHB), indicating the presence of strong electron correlations in this material system~\cite{Evtushinsky2016,Watson2017}. The incoherent feature has also been observed by our ARPES measurements as well as theoretical calculations (see Supplementary Note 6).

Further connection between our ARPES report of OSMP and magnetic excitations via inelastic neutron scattering can also be made. It has been reported that along both the temperature axis~\cite{Xu2014} and substitution axis~\cite{Christianson2013}, low energy magnetic excitations have been observed to change from a commensurate U-shape to a double-rod shape in the absence of any symmetry-breaking orders~\cite{Li2021}. The trend of such behavior is consistent with the observation of the modification of the low energy electronic states along both axes as we have reported, suggesting a common origin.

An OSMP is depicted by the coexistence of localized and itinerant orbitals within one material system~\cite{Anisimov2002,Mukherjee2016,Qiao2017}. Across the FeTe$_{1-x}$Se$_x$ phase diagram, we observe a large dynamic range of the relative coherence of the localized $d_{xy}$ orbital to that of the other itinerant Fe 3$d$ orbitals. In our work, we have discovered a surprisingly transparent way for the $d_{z^2}$ electrons to serve as a diagnostic for the selective localization of the $d_{xy}$ electron states as the OSMP emerges (Fig.~\ref{fig:Fig5}b). To achieve an OSMP in FeTe$_{1-x}$Se$_x$, besides the strong electron-electron interaction, three other ingredients are necessary: i) Crystal field splitting that separates the $d_{xy}$ orbital from $d_{xz}$/$d_{yz}$; ii) Bare bandwidth of the $d_{xy}$ orbital being narrower than that of the $d_{xz}$/$d_{yz}$ orbitals; iii) Suppression of the inter-orbital interactions due to Hund’s $J$~\cite{Si2016}. Our systematic measurement of the fermiology across the FeTe$_{1-x}$Se$_x$ phase diagram establishes the first example to the best of our knowledge amongst the iron-based superconductors where the low energy electronic states are drastically modified due to electron correlation effects, establishing this most correlated FeSC family as an anchoring point between the more itinerant iron pnictides and the more localized cuprates. More generally, the correlation-driven electron reconstruction implies that the Fe 3$d$-electrons are at the boundary between delocalization and localization in an orbital-selective way. As such, the delocalization-localization physics anchors the orbital-selective correlations that have been observed in essentially all the iron-chalcogenide superconductors, including the monolayer FeSe on SrTiO$_3$ that holds the record of superconducting $T_{\rm c}$ in the FeSCs, and in many iron pnictide superconductors~\cite{DeMedici2014,Yi2015}. Consequently, the electrons’  localization-delocalization tendency, along with their proximity to electronic orders, is important both to generating the unusual properties of the normal state and to driving the high-temperature superconductivity.

\section{Methods}
\subsection{Sample growth and characterization}
Our high quality Fe$_y$Te$_{1-x}$Se$_x$ single crystal series were synthesized using the flux method~\cite{Liu2009}. The FeSe single crystal was synthesized using chemical vapor transport~\cite{Chen2019}. The excess Fe in all as-grown Fe$_y$Te$_{1-x}$Se$_x$ single crystals was reduced by annealing in a Te-vapor atmosphere~\cite{Xu2018, Koshika2013}. Crystals with the following Se content were used: \textit{x} = 0, 0.11, 0.19, 0.28, 0.44. The Fe contents of the cleaved surface of these crystals were measured via energy dispersive x-ray spectroscopy following the ARPES measurements and determined to be \textit{y} = 1.08, 1.07, 1.03, 1.01, and 1.00, respectively. 

\subsection{ARPES measurements}
ARPES experiments were performed at beamlines 5-4 and 5-2 of the Stanford Synchrotron Radiation Lightsource with R4000 and DA30 electron analyzers, respectively, beamlines 4.0.3 and 10.0.1 of the Advanced Light Source with R8000 and R4000 electron analyzers, respectively, and the QMSC beamline of the Canadian Light Source with an R4000 electron analyzer. Results were reproduced at all facilities. The angular resolution was set to 0.3$^\circ$ or better. The total energy resolution was set to 15 meV or better. All samples were cleaved \textit{in-situ} below 15 K and all the measurements were conducted in ultra-high vacuum with a base pressure lower than 5 $\times$ 10$^{-11}$ Torr using 72 eV photons unless otherwise noted.

\section{Acknowledgments}
We are thankful to Yu He for enlightening discussions. Parts of the research were performed at the Advanced Light Source and the Stanford Synchrotron Radiation Lightsource, both operated by the Office of Basic Energy Sciences (BES), U.S. Department of Energy (DOE). Work at University of California, Berkeley and Lawrence Berkeley National Laboratory was funded by the U.S. Department of Energy, Office of Science, Office of Basic Energy Sciences, Materials Sciences and Engineering Division under Contract No. DE-AC02-05-CH11231 within the Quantum Materials Program (KC2202) and the Office of Basic Energy Sciences. Part of the research described in this work was performed at the Canadian Light Source, a national research facility of the University of Saskatchewan, which is supported by Canada Foundation for Innovation (CFI), the Natural Sciences and Engineering Research Council of Canada (NSERC), the National Research Council (NRC), the Canadian Institutes of Health Research (CIHR), the Government of Saskatchewan, and the University of Saskatchewan. The ARPES work at Rice University was supported by the Robert A. Welch Foundation Grant No. C-2024, the Alfred P. Sloan Foundation, the Gordon and Betty Moore Foundation's EPiQS Initiative through grant No. GBMF9470, and the US DOE, under Award No. DE-SC0021421. The materials synthesis efforts at Rice are supported by the US DOE, BES, under Contract No. DE-SC0012311 and the Robert A. Welch Foundation, Grant No. C-1839 (P.D.). Theory work at Rice University is supported by the U.S. DOE, Office of Science, BES, under Award No. DE-SC0018197, and by the Robert A. Welch Foundation Grant No. C-1411. Theory work at Renmin University has in part been supported by the National Science Foundation of China Grant No. 11674392, Ministry of Science and Technology of China, National Program on Key Research Project Grant No.2016YFA0300504 and Research Funds of Renmin University of China Grant No. 18XNLG24 (R.Y.). Theory work at Los Alamos was carried out under the auspices of the U.S. Department of Energy National Nuclear Security Administration under Contract No. 89233218CNA000001, and was supported by the LANL LDRD Program. Work at the University of Washington was supported by NSF MRSEC at UW (DMR1719797). Work at Sun Yat-Sen University was supported by the National Natural Science Foundation of China (Grant No. 11904414), Natural Science Foundation of Guangdong (No. 2018A030313055), National Key Research and Development Program of China (No. 2019YFA0705700), and Young Zhujiang Scholar program. Work at Brookhaven is supported by the Office of BES, Division of Materials Sciences and Engineering, U.S. DOE under contract Nos. DE-AC02-98CH10886 and DE-SC00112704. This research was undertaken thanks, in part, to funding from the Max Planck-UBC-UTokyo Center for Quantum Materials and the Canada First Research Excellence Fund, Quantum Materials and Future Technologies Program. 

\section{Author contributions}
M.Y. and R.J.B. proposed and designed the research. M.Y., J.H., M.W., J.S.O., and H.W. carried out the ARPES measurements with the help of D.L., M.H., J.D.D., S.-K.M., M.M., T.M.P., S.G., S.Z. and A.D. The ARPES data were analyzed by J.H. and M.Y. Single crystals of FeTe$_{1-x}$Se$_x$ were synthesized by Z.X. and G.G. FeSe single crystals were synthesized by T.C. under the guidance of P.D. Theoretical calculations were carried out by R.Y., J.-X.Z. and Q.S. Hall measurements were carried out by Q.J. and J.-H.C. J.H. and M.Y. wrote the paper with input from all co-authors. M.Y. and R.J.B. oversaw the project.

\section{Data Availability}
All data needed to evaluate the conclusions are present in the paper and supplementary materials. Additional data are available from the corresponding author on reasonable request.

\section{Competing interests}
The authors declare no competing interests. Ming Yi is an Editorial Board Member for Communications Physics, but was not involved in the editorial review of, or the decision to publish this article.

\newpage 

\begin{figure*}
\includegraphics[width=0.95\textwidth]{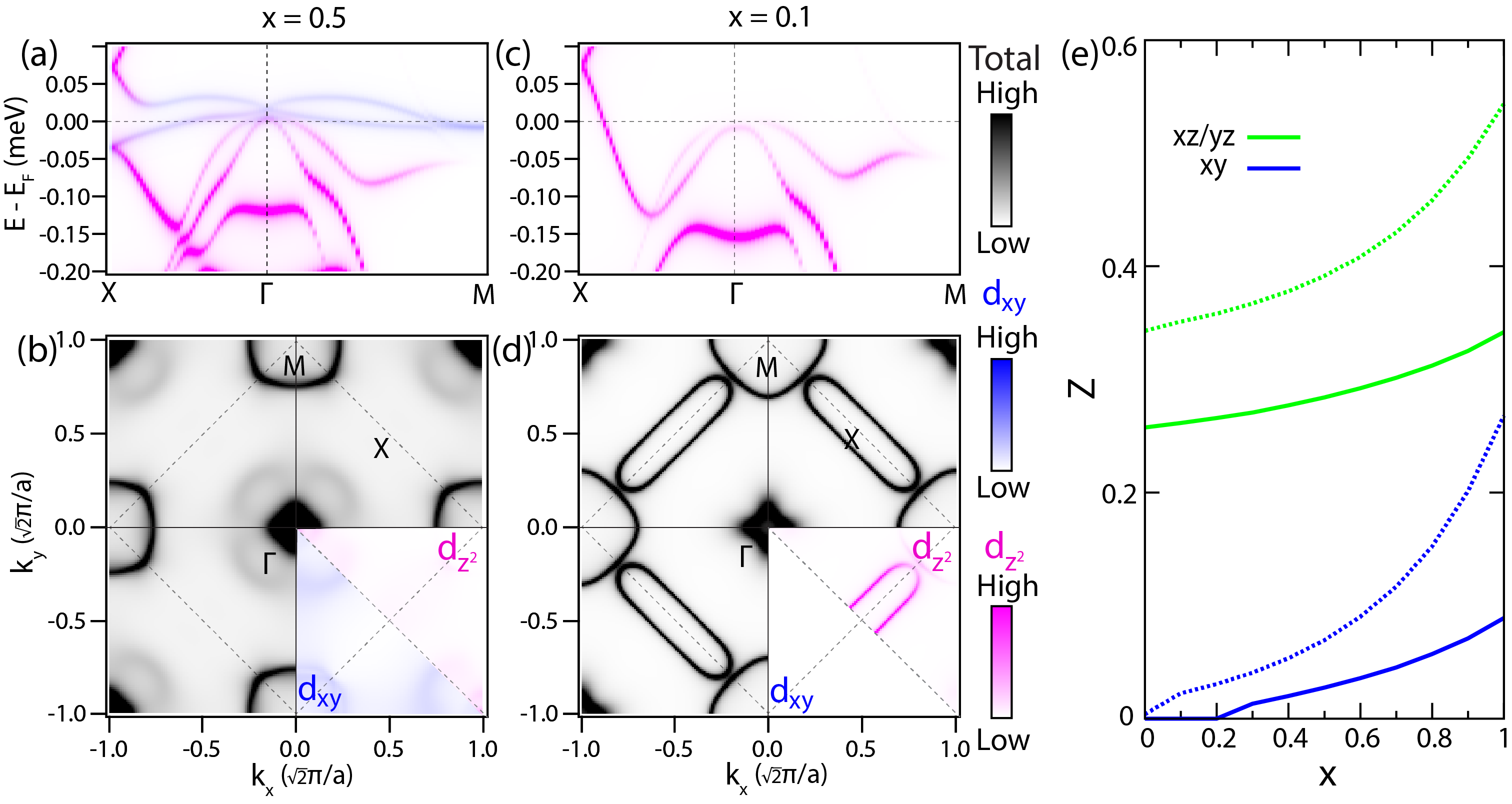}
\caption{\label{fig:Fig4}Theoretical calculations of FeTe$_{1-x}$Se$_x$. (a) Calculated band structure projected onto different orbitals for the strongly correlated metallic phase in FeTe$_{0.5}$Se$_{0.5}$. The magenta intensity indicates $d_{z^2}$ orbital character, whereas blue intensity the $d_{xy}$ orbital. (b) Calculated Fermi surface for FeTe$_{0.5}$Se$_{0.5}$, where the lower right quadrant shows the $d_{z^2}$ and \dxy~intensity respectively. (c)-(d) Same as (a)-(b) but for the orbital-selective Mott phase FeTe$_{0.9}$Se$_{0.1}$. (e) Orbital-resolved coherence factor $Z$ as a function of \textit{x}. The solid line represents $U$ = 3 eV and dashed line represents $U$ = 2.65 eV. $U$ denotes the intraorbital Coulomb repulsion.}
\end{figure*}

\begin{figure}
\includegraphics[width=0.8\textwidth]{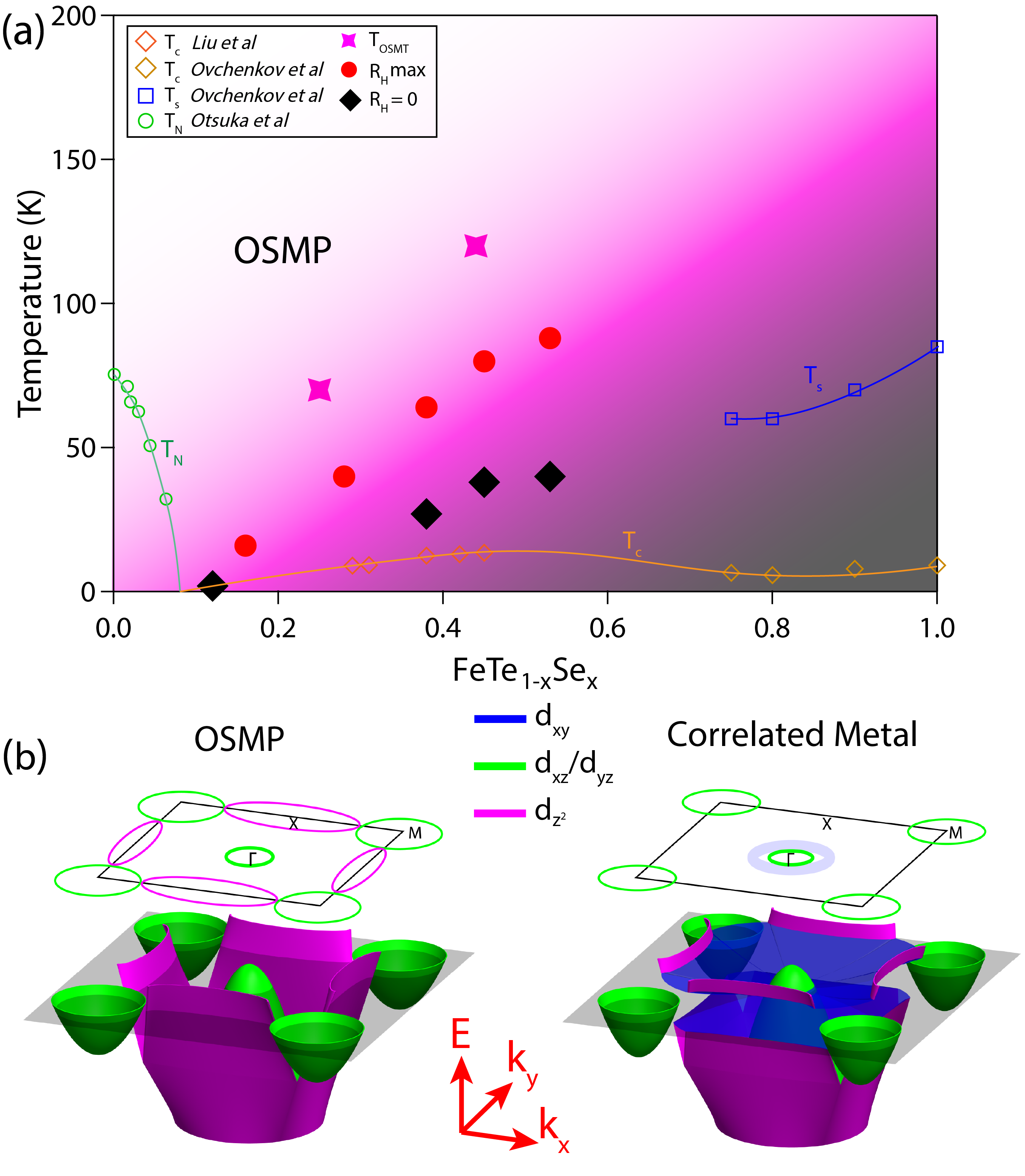}
\caption{\label{fig:Fig5}Phase diagram of FeTe$_{1-x}$Se$_x$. (a) $T_{\rm N}$~\cite{Otsuka2019}, $T_{\rm s}$~\cite{Ovchenkov2019} and $T_{\rm c}$~\cite{Liu2010,Ovchenkov2019} represent the bi-collinear spin-density wave transition, tetragonal to orthorhombic structure transition, and superconducting transition adapted from previous reports. The temperatures for Hall resistivity ($R_{\rm H}$) maximum positions and $R_{\rm H}$ = 0 are extracted from the Hall resistivity measurements~\cite{Jiang2020}. $T_{\rm OSMT}$ is the temperature at which the photoemission spectral weight of $d_{xy}$ orbital is observed to vanish (Supplementary Note 5). OSMT represents orbital-selective Mott transition. The background with gradient color suggests the orbital-selective Mott phase (OSMP) crossover. (b) Schematics showing the key features of the band structure of FeTe$_{1-x}$Se$_x$ in the OSMP and correlated metal phase respectively. The disappearance of band hybridization due to the complete incoherence of $d_{xy}$ orbital is responsible for the Fermi surface reconstruction. }
\end{figure}

\end{document}